\documentclass[aps,pra,reprint,superaddress,showpacs]{revtex4-1}
\usepackage{hyperref}

\usepackage{soul,color}
\usepackage{amsmath}
\usepackage{graphicx}
\usepackage[raggedright,hang,nooneline]{subfigure}
\usepackage{float}

\begin{document}


\title{Parametric Feedback Cooling of Rigid Body Nanodumbbells in Levitated Optomechanics}


\author{T. Seberson$^{1}$ and F. Robicheaux$^{1,2}$}
\affiliation{$^{1}$Department of Physics and Astronomy, Purdue University, West Lafayette, Indiana 47907, USA}
\affiliation{$^{2}$Purdue Quantum Center, Purdue University, West Lafayette, Indiana 47907, USA}

\date{\today}

\begin{abstract}
We theoretically investigate the rigid body dynamics of an optically levitated nanodumbbell under parametric feedback cooling and provide a simplified model for describing the motion. Differing from previous studies, the spin of the nanoparticle about its symmetry axis is considered non-negligible. Simulations reveal that standard parametric feedback cooling can extract energy from two of the five rotational degrees of freedom when the nanoparticle is levitated using a linearly polarized laser beam. The dynamics after feedback cooling are characterized by a normal mode describing precession about the laser polarization axis together with spin about the nanoparticle's symmetry axis. Cooling the remaining mode requires an asymmetry in the two librational frequencies associated with motion about the polarization axis as well as information about the two frequencies of rotation about the polarization axis. Introducing an asymmetric potential allows full cooling of the librational coordinates if the frequencies of both are used in the feedback modulation and is an avenue for entering the librational quantum regime. The asymmetry in the potential needs to be large enough for practical cooling times as the cooling rate of the system depends non-linearly on the degree of asymmetry, a condition that is easily achieved experimentally. 
\end{abstract}


\maketitle

\section{\label{Intro}Introduction}
Optically levitated mesoscopic particles \cite{Vamivakas:16} are known to be ultrasensitive detectors of force, torque, and charge \cite{Ranjit_2016, Hoang2016,Kippenberg1172, doi:10.1063/1.4993555} and provide a range of possible applications such as detection of gravitational waves, fractional charges in bulk matter, and the Casimir torque \cite{Aspelmeyer2014,doi:10.1080/00107514.2014.969492,Xu2017,Arvanitaki2013,Moore2014}. One of the next sought-after goals in levitated optomechanics is for a nanoparticle to reach its quantum mechanical ground state. A nanoparticle in the quantum regime allows exploration of fundamental physics phenomena studying the boundary between the classical and quantum worlds, both mechanically and thermodynamically \cite{Romero-Isart2011,Romero-Isart2011a,1367-2630-12-3-033015,Yin2013,1367-2630-18-1-011002,Millen2014,1367-2630-10-9-095020}. Other micromechanical systems in optomechanics such as microchip resonators offer similar applications \cite{ doi:10.1063/1.4789442} and have been able to attain low occupation numbers, even below $n=1$ \cite{ Chan2011, Safavi-Naeini2012} due to their GHz resonance frequencies and strong coupling to light. However, these systems often require cryogenic cooling or phononic band gaps to suppress decoherence and improve quality factors since they are directly coupled to their environment \cite{Gomis-Bresco2014}. Optically levitated nanoparticles are isolated from rigid structures, eliminating this source of decoherence, and can achieve quality factors $Q>10^{9}$ \cite{Liu:17}. 

Much progress has been made in cooling the translational degrees of freedom (DOF) \cite{Frimmer2016,PhysRevLett.109.103603,Vovrosh:17,Li2011, doi:10.1117/12.2275678, Kiesel14180} with a lowest reported occupation number $n=21$ \cite{Jain:16}. Preventing further reduction in the occupation number is the efficiency with which the position can be detected \cite{PhysRevLett.109.103603,Vovrosh:17}. Shot noise on the detector from the trapping laser hinders the efficiency of position detection, and therefore decreases the effectiveness of the feedback cooling mechanism. Increasing the detection efficiency of the scattered light from the nanoparticle would allow more accurate position detection and is necessary to reach the quantum realm \cite{Jain_thesis,2017arXiv170801203Z}. 

An alternative path to the ground state and a tool for torque sensing \cite{Xu2017,Kuhn2017} is accessing control over the rotational DOF \cite{Zhong2017,Li:18,Kuhn:17, Kuhn2017,Reimann2018, Ahn2018, Arita2013}. Whereas translational mode frequencies are typically in the kHz range, librational mode frequencies can be in the MHz range, possibly offering a more accessible ground state \cite{ Hoang2016}. Cooling the nanoparticle through coupling of the translational and rotational modes has been explored both theoretically and experimentally \cite{ Stickler2016a, Liu:17,Arita2013}. Cooling of the librational modes directly has also been proposed using active feedback schemes \cite{ Zhong2017}. However, these models often assume libration as the sole rotational motion. Describing the rotational dynamics in terms of libration exclusively is a good approximation for particle shapes such as nanorods because of the small moment of inertia about its symmetry axis, but this approximation will break down for particles like dumbbells with more nearly equal moments of inertia. 

In this paper, we seek to investigate the intrinsic coupling between the rotational DOF by considering the classical rigid body dynamics of an optically levitated nanodumbbell with and without parametric feedback cooling. As shown below, for symmetric top-like particles, the spin of the nanoparticle about its symmetry axis, at an angular freqency of $\omega_{3}$, couples the two librational coordinates. This coupling results in two precessional modes amounting to a combination of libration and precession about the polarization axis. In the small angle limit, the equations of motion are of the same form as a charged particle in a two-dimensional harmonic oscillator plus a magnetic field. Previous investigations have dismissed the coupling that leads to precession \cite{ Kuhn2017, 2018arXiv180301778S, Stickler2018a, Stickler2016, Zhong2017, Liu:17, Ahn2018, Zhong2016,Hoang2016}, while recently, the existence of precession motion has been observed for anisotropic nanoparticles \cite{2018arXiv180508042R}. For symmetric top-like particles, $\omega_{3}$ is a conserved quantity, a feature which has important implications when considering ground state cooling of the librational motion. Surprisngly, due to the coupling of the librational coordinates, parametric feedback cooling using a linearized beam is only able to cool one of the two precessional modes. Cooling the remaining mode requires a strong frequency difference in the librational coordinates. 

This paper is organized as follows. Section \hyperref[Model]{II} introduces the classical kinetic and potential energy associated with rotations and provides a simplified model describing the motions. In Sec. \hyperref[Measurement]{III} the signals of two common experimental methods to measure the orientation are calculated utilizing an incident Gaussian laser beam. Section \hyperref[Cooling]{IV} investigates the effects of parametric feedback cooling with linear and elliptical polarization. Simulations of power spectral densities of the measured orientation before and after cooling are also presented. Section \mbox{\hyperref[Discussion]{V}} addresses the effects of laser shot noise and gas collisions.

\section{\label{Model}Theoretical Model}
\begin{figure}[t]  
  \centering
    \includegraphics[width=0.45\textwidth]{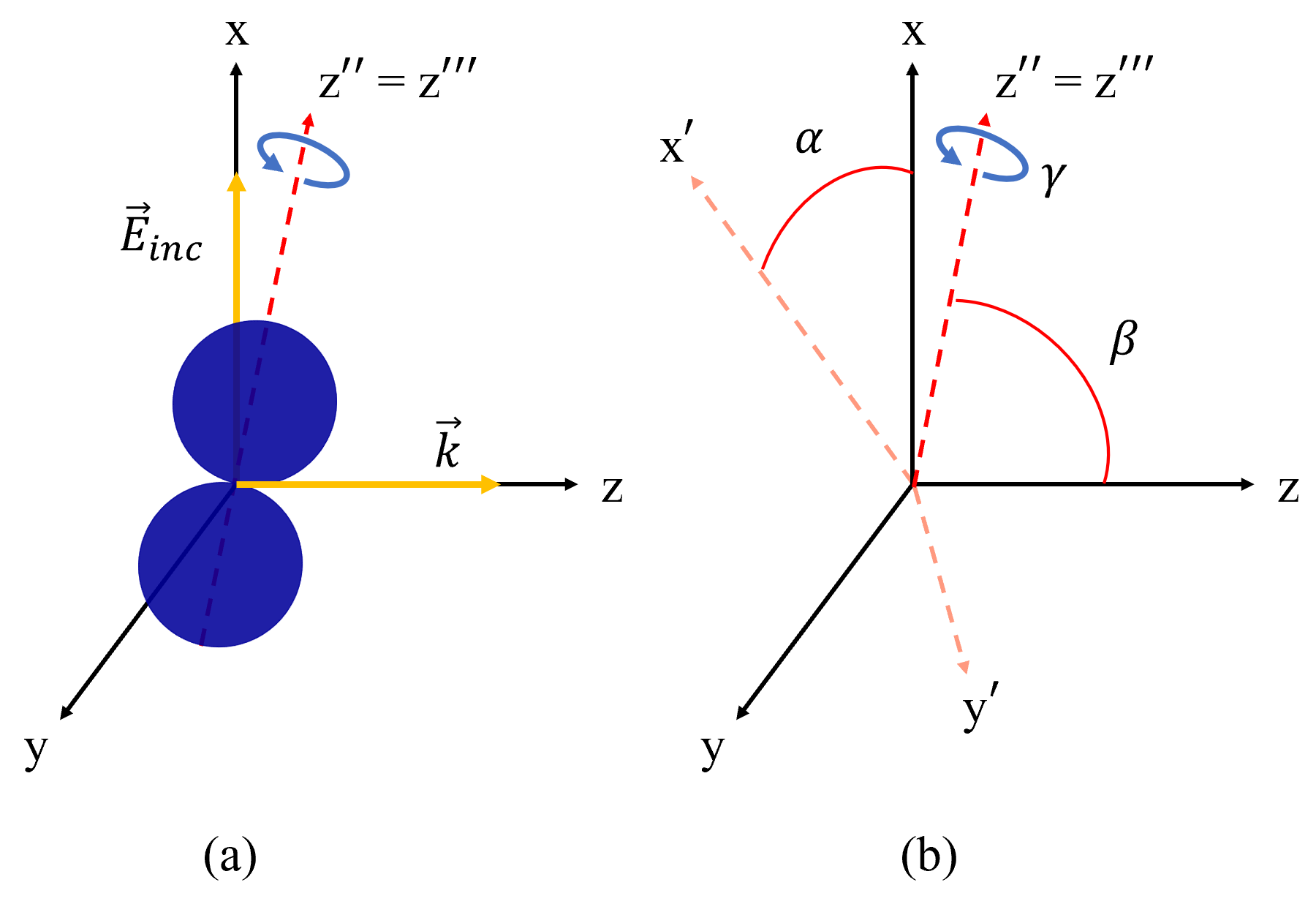}  \label{fig:dumbbell}
     \caption{(a) A nanodumbbell with center of mass confined to the origin is allowed to rotate. The particle has the lowest energy when its long axis ($z'''$-axis) aligns with the laser's electric field polarized in the lab frame $x$-direction. (b) The definition of the Euler angles $\alpha$, $\beta$, $\gamma$ shown in the $z$-$y'$-$z''$ convention. For small angle rotations, the coordinate $\alpha=0+\xi$ describes rotations near the lab frame $x$-axis in the $x$-$y$ plane and $\beta = \pi /2 -\eta$ describes rotations near the lab frame $x$-axis in the $x$-$z$ plane. The coordinate $\gamma$ describes rotations about the $z''=z'''$ axis with $\gamma(t) \approx \omega_{3} t$. For visual clarity, the $x'',y'',x''',y'''$ axes have been omitted from the figure. }
\end{figure}

The system under consideration is a nanodumbbell optically trapped in a laser field. The particle's center of mass is fixed at the origin so that only rotations are considered. The nanodumbbell is composed of two spheres each with mass $M_{s}$ and radius $R$. The spheres are aligned along the $z'''$-axis and touching at the origin, where the triple prime indicates the particle frame coordinate system (see Fig. \hyperref[fig:dumbbell]{1}). It is a symmetric top with principal moments of inertia $I_{x} = I_{y}=\frac{14}{5} M_{s}R^{2}$  and $I_{z} =\frac{4}{5} M_{s}R^{2}$. An amorphous silica nanodumbbell with mass $2M_{s}=1.029\times10^{-17}$ kg, radius $R=85$ nm, $I_{x}=1.041\times10^{-31}~\text{kg}\cdot\text{m}^{2}$, $I_{z}=2.974\times10^{-32}~\text{kg}\cdot\text{m}^{2}$, index of refraction $n=1.458$, and density $\rho=2000~\text{kg}/\text{m}^{3}$ \cite{Ahn2018} is used for the calculations. The laser beam is linearly polarized along the lab frame $x$-direction and propagating in the $z$-direction with a wavelength $\lambda=1550$ nm $\gg R$, power 500 mW, and is focused by a $\text{NA}=0.45$ objective. Because the size of the nanoparticle is much smaller than the wavelength of light, the nanodumbbell is treated as a point dipole with $\vec{E}_{inc}=E_0\hat{x}$, the electric field polarizing the dumbbell, having no spatial dependence. Throughout this paper, the calculations are purely classical, and in what follows, exclude heating from gas collisions and photon scattering. Discussions of the effects due to heating and other noise may be found in Secs. \mbox{\hyperref[Cooling]{IV}} and \mbox{\hyperref[Discussion]{V}}.

The rotational dynamics are governed by the classical equations of motion described by the Euler angles ($\alpha ,\beta ,\gamma$) \cite{SakuraiJ.J.JunJohn2011Mqm,MARION1965361} in the $z$-$y'$-$z''$ convention. To transform from the lab  $(x,y,z)$ frame to the particle $(x''',y''',z''')$ body frame three rotation transformations are made. First, a rotation about the lab frame $z$-axis through an angle $\alpha$ is performed, $(x,y,z)\rightarrow (x',y,'z'=z)$. Then, a rotation about the $y'$-axis is made through an angle $\beta$, $(x',y',z')\rightarrow (x'',y''=y',z'')$. Finally, a rotation about the $z''$-axis is made through an angle $\gamma$, $(x'',y'',z'')\rightarrow (x''',y''',z'''=z'')$. See Appendix \hyperref[appendixA]{A} for further details of the convention used in this paper. 

The kinetic and potential energy are

\begin{align} 
K &= \frac{1}{2}I_{x}(\omega_{1}^2+\omega_{2}^2) + \frac{1}{2}I_{z}\omega_{3}^2 , \\ \begin{split}
U &= -\frac{1}{4} \vec{p}\cdot \vec{E}_{inc} \\
&= -\frac{1}{4}(\alpha_{z}-\alpha_{x})E_{0}^2\cos^2(\alpha)\sin^2(\beta) , \label{eq2}\end{split}
 \end{align}
where  
\begin{align}\begin{split}
\vec{p}&=\overset{\text{\tiny$\leftrightarrow$}}{R}^\dagger \overset{\text{\tiny$\leftrightarrow$}}{\alpha}_{0}  \overset{\text{\tiny$\leftrightarrow$}}{R} \vec{E}_{inc} \\
 &= E_{0}\begin{pmatrix} (\alpha_{z}-\alpha_{x})\cos^2(\alpha)\sin^2(\beta) \\  (\alpha_{z}-\alpha_{x})\sin^2(\beta)\cos(\alpha)\sin(\alpha) \\ (\alpha_{z}-\alpha_{x})\cos(\beta)\sin(\beta)\cos(\alpha) \end{pmatrix} \\
 &\equiv  <p_{x}, p_{y}, p_{z}> , \label{eq3}
\end{split}\end{align} 
is the nanodumbbell polarization vector in the lab frame and $\overset{\text{\tiny$\leftrightarrow$}}{R}$ is the rotation matrix. The $\alpha_{j}$ ($j=x,y,z$) \cite{doi:10.1002/9783527618156.ch5,Trojek:12} are the polarizabilities for an ellipse in the particle frame ($\alpha_{x}=\alpha_{y}$) and are not to be confused with the coordinate $\alpha$. Constant terms in Eqs. (\hyperref[eq2]{2}) and (\hyperref[eq3]{3}) have been omitted as they do not affect the particle's rotational dynamics. The $\omega_{i}$ ($i=1,2,3$) and the full equations of motion may be found in Appendix \hyperref[appendixA]{A}. It should be noted that as a consequence of the nanodumbbell's symmetry, the angular momentum about the nanoparticle's symmetry axis, $I_{z}\omega_{3}$, is a constant of the motion. In this configuration, each Euler angle has an intuitive definition for small amplitude oscillations; $\alpha$ defines libration in the $x$-$y$ plane, $\beta$ defines libration in the $x$-$z$ plane, and $\gamma$ corresponds to angles of rotation about the $z'''$-axis. 

\begin{figure}[H]  
\centering
  \hspace*{-0.5cm}\includegraphics[width=0.45\textwidth]{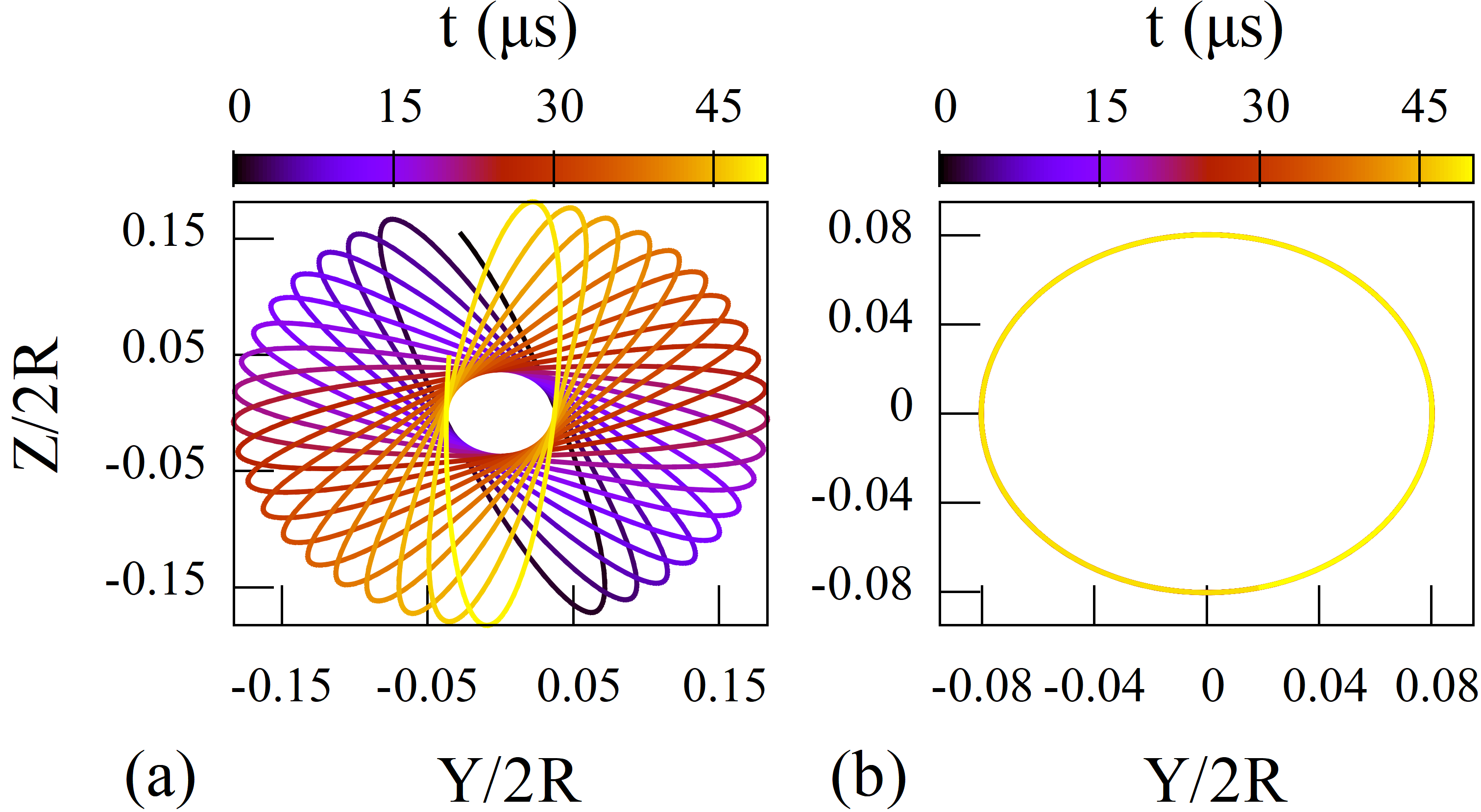}  \label{fig:trajectories}
    \caption{Trajectory of the nanoparticle's $z'''$-axis, projected on the lab frame $y$-$z$ plane found by simulating the full equations of motion and using linear polarization. Here, $Z/2R=\cos(\beta)$, $Y/2R=\sin(\beta)\sin(\alpha)$ define the location of the $z'''$-axis; in the small angle limit $Z/2R\approx \eta$, $Y/2R\approx \xi$. (a) The particle's long axis moves in two joint motions, one describing libration and the other describing precession about the polarization axis. (b) Final trajectory of the long axis after parametric feedback cooling. The motion has reduced to pure precession. }
\end{figure}

The attractive potential, Eq. (\hyperref[eq2]{2}), causes the particle to oscillate about the polarization axis in two joint motions (see Fig. \hyperref[fig:trajectories]{2(a)}). The two motions are most easily seen under a small angle approximation. It is energetically favorable for the particle's long axis ($z'''$-axis) to align with the electric field and is therefore localized near the lab frame $x$-axis. This corresponds to $\alpha$ nearing towards zero or $\pi$, and $\beta$ near $\pi/2$. Allowing the two coordinates to make small oscillations about the $x$-axis, $\alpha \rightarrow 0 + \xi$ , $\beta \rightarrow \frac{\pi }{2} - \eta$, with $\xi ,\eta$ small, the equations of motion to first order become 
 \begin{align} \begin{split}
\ddot{\xi} &= \Big[-\frac{\omega^{2}}{2}\sin(2\xi) - \omega_{c}\dot{\eta}\sec(\eta) + 2\dot{\eta}\dot{\xi}\tan(\eta)\Big] \\
&\approx -\omega^2\xi - \omega_{c}\dot{\eta} , \label{eq4} \end{split}\\ \begin{split}
\ddot{\eta}  &= \cos(\eta)\Big[ -\omega^{2}\sin(\eta)\cos^{2}(\xi) + \omega_{c}\dot{\xi} - \dot{\xi}^{2}\sin(\eta)\Big] \\
&\approx -\omega^2\eta + \omega_{c}\dot{\xi} , \label{eq5} \end{split}
\end{align}
where $\omega^{2} =\frac{1}{2}(\alpha_{z}-\alpha_{x})E_{0}^{2}/I_{x} $ and $\omega_{c}=\left(I_{z}/I_{x}\right)\omega_{3}$. The first term on the right hand side of Eqs. (\ref{eq4}) and (\ref{eq5}) amounts to libration about the polarization axis due to the trapping potential which has been seen before in \cite{Zhong2017,2018arXiv180301778S}. The second term containing  $\omega_{c}$ couples the four DOF and is responsible for precession about the $x$-axis. The precession is a consequence of the non-zero angular momentum about the symmetry axis, $I_{z}\omega_{3}$. Precession has recently been seen for anisotropic particles in an elliptically polarized beam \cite{2018arXiv180508042R}, but with $\alpha$ precessing around the lab frame $z$-axis with $\beta$ roughly fixed. As in the case for thin nanorods, the motion reduces to pure libration in the limit $I_{z}\rightarrow 0$. The equation of motion for $\gamma$ is not directly affected by the potential and largely evolves with time as $\gamma(t) \approx \omega_{3} t$ in the small angle approximation (see Appendix \ref{appendixA} Eq. (\hyperref[eqA12]{A12})). 

The transformation of the $z'''$-axis into the lab frame, $\hat{r}_{z'''}=\overset{\text{\tiny$\leftrightarrow$}}{R}^\dagger \hat{z}'''$, determines the location of the tip in the ($x$, $y$, $z$) coordinate system,
\begin{align}
\hat{r}_{z'''} = \begin{pmatrix} \sin(\beta) \cos(\alpha)  \\  \sin(\beta)\sin(\alpha)  \\ \cos(\beta)  \end{pmatrix} 
 \approx \begin{pmatrix} 1  \\  \xi  \\ \eta  \end{pmatrix} ,
\end{align}
where in the last step the small angle approximation was made. It is seen that $\xi$ and $\eta$ play the role of the $y$ and $z$ coordinates defining the location of the tip. By introducing a vector that specifies the projection of the $z'''$-axis on the $y$-$z$ plane, $\vec{\rho}_{z'''}=<0,\xi,\eta>$, it is possible to combine Eqs. (\ref{eq4}), (\ref{eq5}),
\begin{equation}
\ddot{\vec{\rho}}_{z'''} = -\omega^{2}\vec{\rho}_{z'''} - \dot{\vec{\rho}}_{z'''}\times \vec{\omega}_{c}, \label{eq7}
\end{equation}
where $\vec{\omega}_{c}=\omega_{c}\hat{x}=\left(I_{z}/I_{x}\right)\omega_{3}\hat{x}$. The last term in Eq. (\ref{eq7}) has the familiar form of the force on a charged particle in a magnetic field. The two joint motions now become clear as a combination of harmonic oscillations in a static pseudo-magnetic field. Thus, as long as $\omega_{c}$ is non-zero, the full dynamics of the nanoparticle must be described as a combination of libration and precession, as opposed to just libration. For a nanodumbbell at room temperature, $T=300$ K, the average value of $\omega_{c} \sim \sqrt{k_{B}TI_{z}}/I_{x}\sim 10$ kHz, where $k_{B}$ is the Boltzmann constant. While $\omega \sim 100~\text{kHz}-1~\text{MHz}\gg\omega_{c}$, the coupling that results due to the $\sim$10 kHz frequency is a resolvable feature in the power spectral density and is a non-negligible effect when considering parametric feedback cooling, as will be discussed in Sec. \ref{Cooling}. 

The librational frequency $\omega$ scales with the radius as $\omega^{2}\sim 1/R^{2}$ suggesting that a particle of smaller size is beneficial for ground state cooling. However, the polarizability and moment of inertia scale as $\alpha_{j}\sim R^{3}$, $I_{j}\sim R^{5}$ implying that the particle will be less confined and more unstable in the optical trap as the size decreases. In effect, a smaller radius will be more likely to escape the trap and will produce a broader power spectral density. Further, $\omega_{c}\sim 1/R^{5/2}$ showing that as the size of the particle decreases the precessional phenomenon is more pronounced. 

Equations (\ref{eq4}) and (\ref{eq5}) admit two normal modes, 
\begin{align}
\xi(t) &= A_{+} \cos(\omega_+ t +\delta_{+}) + A_{-} \cos(\omega_{-} t +\delta_{-})  , \label{eq8} \\
\eta(t) &= A_{+} \sin(\omega_+ t +\delta_{+}) - A_{-} \sin(\omega_{-} t +\delta_{-})  , \label{eq9}
\end{align}
with $\omega_{\pm} = \frac{1}{2}\left( \Omega \pm\omega_{c} \right)$, $\Omega=\sqrt{4\omega^2+\omega_c^2}$, and the $A_{\pm}$, $\delta_{\pm}$  determined by initial conditions. Each mode circles the polarization axis at a particular frequency with the (+) mode advancing clockwise and the ($-$) mode counterclockwise. The superposition of the two modes results in the libration and precession mentioned above. Thus, as will be discussed in Sec. \ref{exp_sig}, the power spectral density of $\xi$ or $\eta$ should exhibit two peaks at $\omega_{\pm}$.

Since the coordinates $\alpha$ and $\beta$ completely describe the location of the nanodumbbell's tip projected on the lab frame axes, it is possible to track the rotational evolution about the polarization axis while simulating the full equations of motion. Figure \hyperref[fig:trajectories]{2(a)} plots the $z'''$-axis projection on the lab frame $z$-$y$ plane (i.e. $Z/2R=\cos(\beta)$, $Y/2R=\sin(\beta)\sin(\alpha)$ is plotted versus time. Note that in the small angle limit $Z/2R\approx \eta$, $Y/2R\approx \xi$). The particle's tip undergoes fast oscillations enveloped in a slower precession motion about the $x$-axis, qualitatively consistent with the dynamics seen in the small angle approximation. 

\section{\label{Measurement}Measuring the Orientation}
It is possible to determine the orientation of the nanoparticle with respect to $\alpha$ and $\beta$ through different types of measurements. A common method of measuring $\alpha$ libration \cite{Reimann2018,Ahn2018,Kuhn:17,Arita2013} is to first send the forward scattered light from the nanoparticle and the laser beam through a 45$^\circ$ polarized beamsplitter (PBS). The light exits the PBS in two different directions with orthogonal polarizations. A measurement is obtained by reading the signal of each polarization state on a photodetector and taking the difference between the two signals. To determine what is measured in this procedure, consider a Gaussian laser beam incident on the dumbbell,
\begin{equation}
\vec{E}_{inc} = E_{0} \frac{\omega_{0}}{\omega(z)} e^{\frac{-\rho^2}{\omega^2(z)}} e^{i(kz+\frac{k\rho^2}{2R(z)} -\psi(z))}\hat{x}. \label{eq10}
\end{equation}
The Gaussian beam is defined with $\omega_{0}$ the beam waist, $\omega(z)=\omega_{0}\sqrt{1+(z/z_{R})^{2}}$ with $z_{R}=\pi\omega^{2}_{0}/\lambda$ the Rayleigh range, $\rho^{2}=x^{2}+y^{2}$, $k=2\pi/\lambda$, $R(z)=z[1+(z_{R}/z)^{2}]$, and $\psi(z)=\arctan(z/z_R)$. The scattered light is determined by the electric and magnetic fields for a dipole in the far field \cite{jackson_classical_1999}
\begin{align}
\vec{H} &= \frac{ck^2}{4\pi}(\hat{r}\times\vec{p})\frac{e^{ikr}}{r} , \\ 
\vec{E} &= Z_{0} \vec{H}\times\hat{r} ,
\end{align}
where $\vec{r}$ is in the direction of observation, $c$ the speed of light, and $Z_{0}$ the impedance of free space. After exiting a collimating lens \cite{novotny_hecht_2006}, the light is split by a 45$^\circ$ PBS. The transverse components of the electric field exiting the PBS are
\begin{align}
\vec{E}_{+}(x,y)&= \frac{1}{\sqrt{2}} \Big(E_x(x,y)+E_y(x,y) \Big)\hat{e}_{+} ,\\
\vec{E}_{-}(x,y)&= \frac{1}{\sqrt{2}} \Big(E_y(x,y)-E_x(x,y) \Big)\hat{e}_{-} ,
\end{align}
where $E_{x,y}$ are the $x$ and $y$ components of the total electric field following the collimating lens and $\hat{e}_{\pm}$ designate the two split polarization states after the PBS. The magnetic field undergoes a similar transformation. A measurement is performed by taking the difference between the two signals measured at their respective detectors
\begin{equation}
P_{45^\circ}=\int_{-\infty}^{\infty}\int_{-\infty}^{\infty} \left(\vec{S}_{+}\cdot\hat{z} - \vec{S}_{-}\cdot\hat{z} \right)dydx ,
\end{equation}
where $\vec{S}_{\pm} = \frac{1}{2}Re[\vec{E}_{\pm}\times\vec{H}^*_{\pm}]$ is the Poynting vector. Performing the integration gives a homodyne term that is proportional to the $y$-component of the polarizability from Eq. (\ref{eq3})
\begin{equation}
P_{45^\circ} \propto p_y \propto \sin^2(\beta)\cos(\alpha)\sin(\alpha). \label{eq16}
\end{equation}
Considering that $\alpha \rightarrow 0 + \xi$ and $\beta \rightarrow \frac{\pi }{2} - \eta$, 
\begin{equation}
p_y\propto \cos^2(\eta)\cos(\xi)\sin(\xi) \approx \xi, \label{eq17}
\end{equation}
which is the angle describing the extent to which the nanoparticle's long axis has deviated from the polarization axis in the $x$-$y$ plane. 
Following the same procedure above, after the collimating lens a split detection measurement is performed (which is often used to track the transverse translational motion \cite{Frimmer2016,PhysRevLett.109.103603,Vovrosh:17,Li2011, doi:10.1117/12.2275678, Kiesel14180}) and the homodyne term is examined
\begin{align}\begin{split}
P_{x}&=\int_{0}^{\infty}\int_{-\infty}^{\infty} \left(\vec{S}\cdot\hat{z}\right)dydx -\int_{-\infty}^{0}\int_{-\infty}^{\infty} \left(\vec{S}\cdot\hat{z} \right)dydx \\
& \\
  &\propto p_z \propto \cos(\beta)\sin(\beta)\cos(\alpha) \approx \eta ,
\end{split}\end{align}
which is the angle describing the extent to which the nanoparticle's long axis has deviated from the polarization axis in the $x$-$z$ plane. Thus, both angles that will be required for parametric feedback cooling in the next section can be detected. Note that it is not possible to measure $\gamma$ directly using these methods since it is not contained within the polarization vector in Eq. (\hyperref[eq3]{3}). 

Whereas the detection of translational motion relies on the $\pi/2$ Gouy phase shift from $\psi(z)$ in Eq. (\hyperref[eq10]{10}) \cite{doi:10.1063/1.5008396}, for a nanoparticle centered at the origin, the Gouy phase shift hinders detection of rotational motion. The homodyne terms in the Poynting vector evaluated far from the nanoparticle are left purely imaginary and require the imaginary part of the polarizability for orientational detection. The imaginary part of the polarizability is usually smaller than the real part \cite{Dholakia2010,Trojek:12} for the types of particles used in levitated optomechanics and is two orders of magnitude smaller for a $R=85$ nm amorphous silica dumbbell in a $\lambda = 1550$ nm laser field. 

However, the signal may become real, and therefore larger in magnitude, if the particle is not centered at the origin, but pushed away from the focus by the laser beam in the axial direction. In Appendix \hyperref[appendixB]{B} the size of the displacement in the $z$-direction is estimated by considering the radiation pressure on a nanodumbbell in the point dipole limit. Relative to the Rayleigh range, $z_{R}$, the displacement $z_{d}$ is
\begin{equation}
\frac{z_{d}}{z_{R}} = \Big(\frac{32\overline{\alpha}}{3}\Big)\Big(\frac{\pi R}{\lambda}\Big)^{3}\Big(\frac{1}{\text{NA}}\Big)^{2} , \label{eq19}
\end{equation}
where $\overline{\alpha}$ is a unitless parameter defined in the polarizability as $\alpha_{0} = 4\pi \epsilon_{0} R^{3}\overline{\alpha} $. For $\overline{\alpha} = 0.59$, $\lambda = 1550$ nm, $\text{NA} = 0.45$, $R=85$ nm, $z_{d}/z_{R}\sim 0.14$. This ratio becomes important for measurements as $e^{iz_{d}/z_{R}}\sim (1+iz_{d}/z_{R})$ is a prefactor in the polarizability matrix when considering $\psi(z)$ in Eq. (\ref{eq10}), effectively reducing the measured signal by this ratio. 

\section{\label{Cooling}Parametric feedback cooling}

Parametric feedback cooling utilizes a laser beam to trap a particle and cool its motion simultaneously through modulation of the laser power at twice the particle's oscillation frequency\mbox{ \cite{PhysRevLett.109.103603}}. To obtain a signal at twice the oscillation frequency, the coordinate to be cooled is multiplied by its time derivative $q\dot{q}$, for an arbitrary coordinate $q$. 

The analyses here assume perfect and instantaneous measurements of $q\dot{q}$, which cannot be achieved in practice. Shot noise heating due to photon scattering and the effects of gas collisions are also not included to simplify the analysis. While the heating mechanisms do determine the lowest energy attainable for a fixed cooling power, the dynamical effects of gas collisions do not become important for pressures $\sim10^{-3}$ Torr or lower and photon shot noise does not become important until the heating rate is near the cooling rate. See Sec. \mbox{\hyperref[Discussion]{V}} for further discussion of these effects and the inclusion of noise. The results that follow thus provide a fundamental limit to cooling, irrespective of the limitations set by quantum mechanics or practical experimental parameters such as orientation detection efficiency. 

It is also worth mentioning that since the equations of motion for $\gamma$ are unaffected by the trapping potential, with the dynamics determined largely by the conserved $\omega_{3}$, parametric feedback cooling only directly affects the $\alpha,\dot{\alpha}$ and $\beta,\dot{\beta}$ DOF. For this reason, the focus will be on the motions associated with $\alpha$ and $\beta$ ($\xi$ and $\eta$), as it is not possible to cool the nanoparticle's spin about its symmetry axis using parametric feedback cooling.

\subsection{\label{Linear_cooll}Linear Polarization}

\begin{figure*}[t] 
  \centering
	 \includegraphics[width=1. \textwidth]{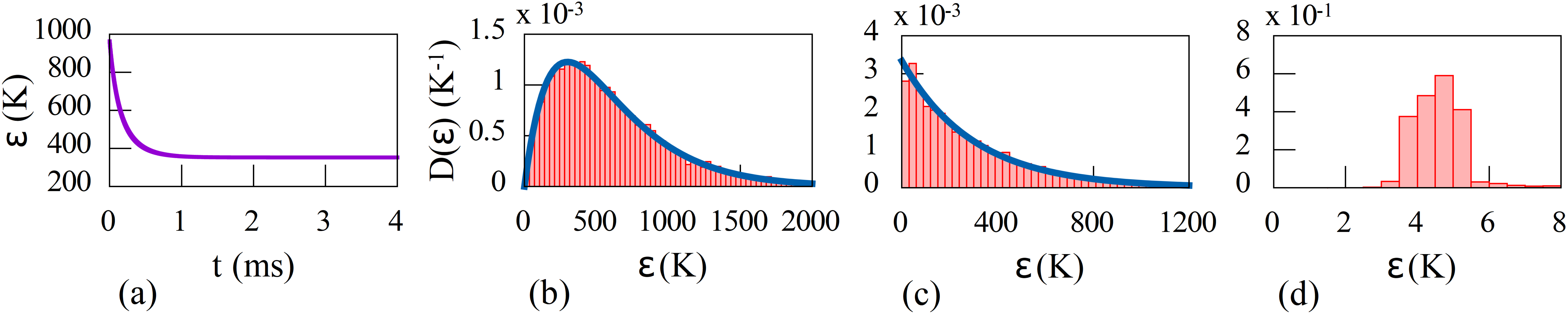}  \label{fig:histograms} 
    \caption{(a) Energy of the four degrees of freedom versus time under parametric feedback cooling during a single trajectory. The energy plateaus to a non-zero value as the nanoparticle circles the polarization axis. (b) Distribution of initial energies used before parametric feedback cooling. The distribution follows the blue line, a Maxwell-Boltzmann distribution with four degrees of freedom at a temperature of 300 K. (c) Distribution of energies following parametric feedback cooling with linear polarization. The distribution has a mean of 312 K and is similar to the blue line, a Maxwell-Boltzmann distribution with two degrees of freedom at a temperature of 300 K. The final energies are taken as the last data point in runs similar to that of (a). (d) Final energy distribution of the nanoparticle following feedback cooling with elliptical polarization for $\theta = \frac{4\pi}{32}$. Each distribution in this figure is composed of 12,000 runs implemented with a cooling strength $\chi=10^7~\text{s}/\text{m}^{2}$.}
\end{figure*}
                                        
The equations of motion for the dumbbell in the small angle approximation from Sec. \hyperref[Model]{II.} Eqs. (\hyperref[eq4]{4}) (\hyperref[eq5]{5}) under feedback cooling become
\begin{align}
\ddot{\xi} &= -\omega^2(1+\chi R^2q\dot{q})\xi - \omega_{c}\dot{\eta} , \label{eq20} \\
\ddot{\eta}  &= -\omega^2(1+\chi R^2q\dot{q})\eta + \omega_{c}\dot{\xi} , \label{eq21}
\end{align} 
where $\chi$ is the cooling strength that sets the amplitude of the power modulation. Choosing to measure and feedback $q\dot{q}=\xi\dot{\xi}$ into Eqs. (\ref{eq20}), (\ref{eq21}), the average cooling power is calculated as
\begin{align} \begin{split}
<P>_{\xi\dot{\xi}} &= <\frac{dE}{dt}> \\
& = \Big(\frac{I_{x}}{2}\Big) < \frac{d}{dt}\Big[ \dot{\xi}^{2} +  \dot{\eta}^{2} +\omega^{2}\Big(\xi^{2} +\eta^{2} \Big) \Big] > \label{eq22} \\
& = -I_{x}\omega^{2}\chi R^{2}<\xi\dot{\xi}\Big( \eta\dot{\eta} + \xi\dot{\xi}\Big)>.
\end{split} \end{align}
Inserting Eqs. (\ref{eq8}) and (\ref{eq9}) into Eq. (\ref{eq22}) gives the cooling rate in terms of the normal mode amplitudes $A_{\pm}$. Performing the derivatives in Eq. (\ref{eq22}) with the $A_{\pm}$ slow compared to $\omega_{\pm}$ and averaging the sinusoidal factors over one cycle gives 
\begin{equation}
<P>_{\xi\dot{\xi}}=-\frac{1}{4}I_{x}\omega^{2}\chi R^{2}\Omega^2\Big[ A_{+}(t)A_{-}(t)\Big]^2 , \label{eq23} 
\end{equation}
which implies that cooling is effective until one mode is removed from the motion. As $t\rightarrow \infty$ the particle will fully precess about the polarization axis with no libration (see Fig. \hyperref[fig:trajectories]{2(b)}). The result of a single mode remaining is a plateau in the energy over time as shown in Fig. \hyperref[fig:histograms]{3(a)}. Choosing to measure and feedback the frequency $\eta\dot{\eta}$ produces the same result while the addition of the two, $q\dot{q}=\eta\dot{\eta}+\xi\dot{\xi}$, delivers the same effect at twice the rate since $\eta$ and $\xi$ oscillate at the same frequency. 

Using the conserved quantity 
\begin{equation} 
\frac{d}{dt} \Big[\xi\dot{\eta}-\eta\dot{\xi}-\frac{\omega_{c}}{2}(\xi^{2}+\eta^{2})  \Big]=0, \label{eq24}
\end{equation} 
together with Eqs. (\ref{eq8}) and (\ref{eq9}) gives the exact expression $\frac{d}{dt}\Big( A^{2}_{-}(t)-A^{2}_{+}(t)\Big)=0$. This condition shows that as one mode is cooled completely, the second mode ceases to be time dependent, facilitating the notion that there is a limit to how much energy is removed from the motion. 

To investigate the extent of possible cooling, a nanodumbbell is initially prepared with a thermal distribution at $T=300~K$ and several thousand cooling runs are simulated using the full equations of motion. The simulations are run using a fourth order Runge-Kutta adaptive step algorithm \cite{Press:2007:NRE:1403886} with random initial conditions conforming to a Boltzmann distribution. The initial frequencies of rotation are found by
\begin{equation}
\omega_{i} = \sqrt{\frac{k_{B}T}{I_{j}}}dW ,
\end{equation}
with $(i,j)=((1,x) ,( 2,y) ,(3,z))$ and $dW$ a guassian random number with zero mean and unit variance. The initial coordinate values $\alpha , \beta$ are established through rejection sampling of the potential,
\begin{equation}
\mathcal{P}(\alpha,\beta)=e^{-\big(U(\alpha,\beta)-U(0,\frac{\pi}{2})\big)/k_{B}T  } ,
\end{equation}
and $\gamma$ is initialized as a uniformly distributed random number between $0$ and $2\pi$. 

Figure \hyperref[fig:histograms]{3(b)(c)} shows the energy distributions before and after cooling for a cooling strength $\chi=10^7~\text{s}/\text{m}^{2}$ and feedback frequency $p_y\dot{p}_{y}$ (see Eqs. (\ref{eq3}), (\ref{eq16}), and (\ref{eq17})). In the figure, $D(\varepsilon)$ is the probability energy density with $\int_{0}^{\infty}D(\varepsilon)d\varepsilon=1$. As cooling extracts energy from the $\alpha,\dot{\alpha}$ and $\beta,\dot{\beta}$ DOF exclusively, in Fig. \hyperref[fig:histograms]{3} the shifted energy $\varepsilon=E-\frac{1}{2}I_{z}\omega^2_3$ is used where $E=K+U(\alpha,\beta)-U(0,\pi/2)$ is the total energy adjusted so that $0~K$ is the minimum energy. The blue lines in Fig. \hyperref[fig:histograms]{3(b)(c)} are plots of the Maxwell-Boltzmann distribution function $A\varepsilon^{n}\exp(-\varepsilon/300)$ for four ($n=1$) and two ($n=0$) DOF, respectively. As expected, the initial energies follow a Maxwell-Boltzmann distribution with an average energy $602~\text{K}\sim\frac{4}{2}T$ corresponding to four quadratic DOF. In the final energy distribution, it is seen that effectively two DOF have been removed due to cooling; the final energy distribution has a mean energy of $312~\text{K}\sim\frac{2}{2}T$, corresponding to two uncooled quadratic DOF. The two DOF remaining is consistent with the nanoparticle's long axis circulating around the polarization axis at a fixed non-zero angle, qualitatively seen in both the small angle approximation and the full simulation. The result that parametric feedback cooling is unable to cool the nanoparticle's motion completely even if both coordinate frequencies are known is one of the important results of this paper. It is clear that cooling into the quantum regime is not possible utilizing a perfectly linearized beam and standard parametric feedback cooling, even if both angular DOF can be detected. 

\subsection{\label{Elliptical_cooll}Elliptical Polarization}
The issue with the previous section's strategy for cooling is the coupling between $\eta$ and $\xi$ due to the spin about the symmetry axis and their similar frequencies of rotation about the polarization axis. To cool further requires breaking the symmetry between the two DOF responsible for the precession motion. In this section this symmetry is broken by introducing a potential that produces different librational frequencies for the two coordinates $\eta$ and $\xi$. This can be acheived through elliptical polarization, using two perpendicular laser beams incident on the nanoparticle, or general asymmetries found in a focused laser beam's gradient \cite{novotny_hecht_2006}. Here, elliptical polarization is used with $\vec{E}_{inc}=E_{0}<\cos\theta,i\sin\theta,0>$. This alters Eqs. (\hyperref[eq4]{4}) and (\hyperref[eq5]{5}) 
 \begin{align}
\ddot{\xi} &= -\omega^{2}_{\xi}\xi - \omega_{c}\dot{\eta} \label{eq27} , \\
\ddot{\eta}  &= -\omega^{2}_{\eta}\eta + \omega_{c}\dot{\xi} \label{eq28},
\end{align}
where $\omega^{2}_{\xi}=\omega^{2} \big( \cos^{2}\theta-\sin^{2}\theta \big)$ , $\omega^{2}_{\eta}=\omega^{2}\cos^{2}\theta$. The normal modes and further details of this system are described in Appendix \hyperref[appendixC]{C}. Feedback cooling using either $q\dot{q}=\xi\dot{\xi}$ or $q\dot{q}=\eta\dot{\eta}$ gives the following average cooling rates
\begin{equation}\begin{split}
<P&>_{\xi\dot{\xi}} = \\
&\Big[ A_{+}(t)\Big]^{4}y_{1} -\Big[ A_{-}(t)\Big]^{4}y_{2} - \Big[ A_{+}(t)A_{-}(t)\Big]^{2}y_{3} \label{eq29} , \end{split}\end{equation}
\begin{equation} \begin{split}
<P&>_{\eta\dot{\eta}} = \\
-&\Big[ A_{+}(t)\Big]^{4}z_{1} +\Big[ A_{-}(t)\Big]^{4}z_{2} - \Big[ A_{+}(t)A_{-}(t)\Big]^{2}z_{3}\label{eq30} , \end{split}
\end{equation}
where the $y_{i}$, $z_{i}$ ($i=1, 2, 3$) are positive and constant for a fixed electric field strength (see Appendix \hyperref[appendixC]{C}) and reduce to Eq. (\hyperref[eq23]{23}) for $\theta=0$. Equations (\hyperref[eq29]{29}), (\hyperref[eq30]{30}) show a combination of heating and cooling with each choice of feedback frequency having preference of cooling a particular mode. In this arrangement one mode is cooled while the other heats, ultimately leading to heating. Simulations of energy versus time while feeding back the frequency $\eta\dot{\eta}$ or $\xi\dot{\xi}$ show the energy increasing indefinitely, sometimes following an initial period of cooling depending on the initial conditions.

However, feeding back both coordinate's frequencies in the form $q\dot{q}=\eta\dot{\eta}+\xi\dot{\xi}$ will lead to cooling of both modes. The cooling rate $<P>_{\xi\dot{\xi}+\eta\dot{\eta}}=<P>_{\xi\dot{\xi}}+<P>_{\eta\dot{\eta}}$ is negative for all $\omega_{\eta}>\omega_{\xi}>\omega_{c}$ (see Appendix\hyperref[appendixC]{C}) which are the conditions considered in this paper. Figure \hyperref[fig:histograms]{3(d)} shows the final energy distribution with this choice of feedback for a fixed cooling strength $\chi=10^{7}~\text{s}/\text{m}^{2}$ and $\theta=\frac{4\pi}{32}$. The particle's accessible DOF have been cooled significantly compared to the case for linear polarization. 

\begin{figure}[h]  
  \centering
  \includegraphics[width=0.47\textwidth]{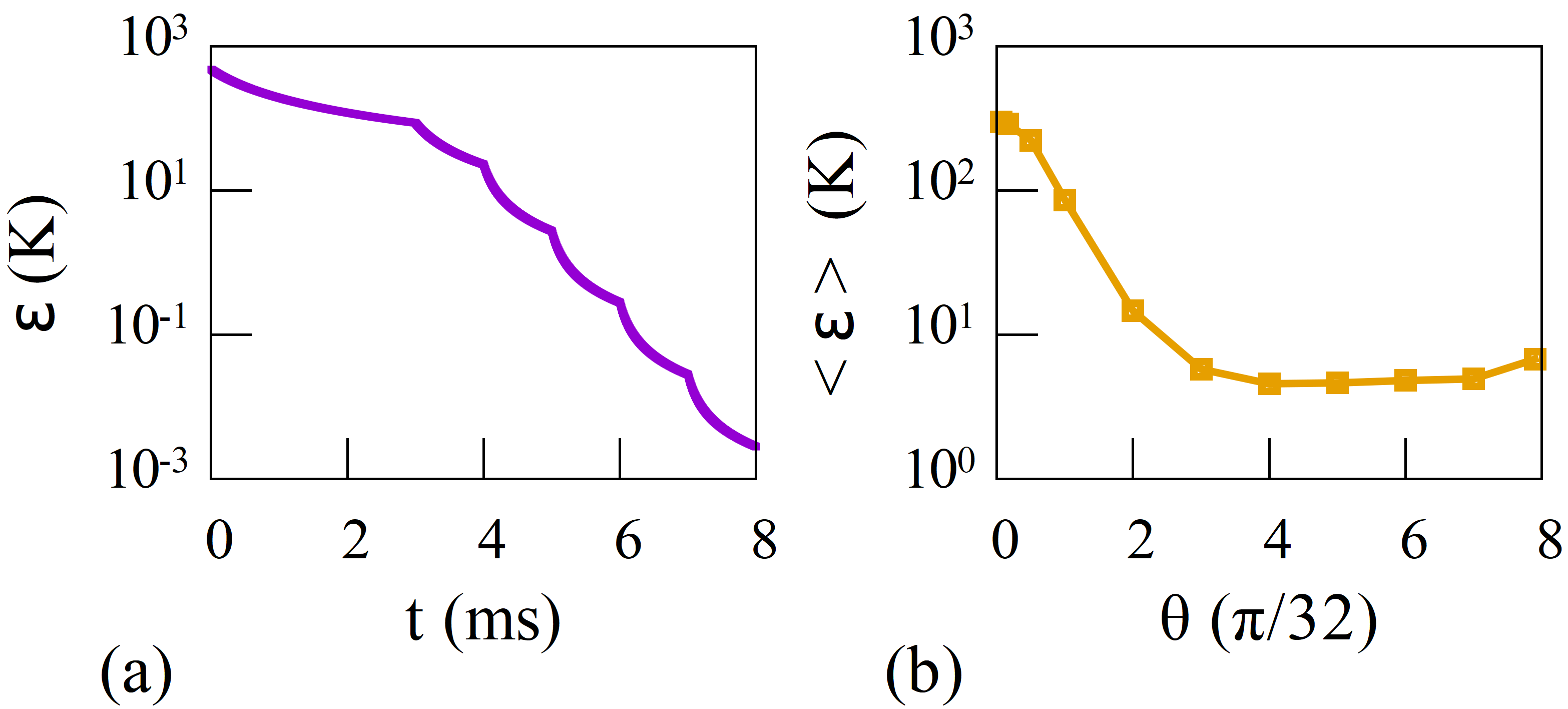}  \label{fig:power_depend}
    \caption{Plots showing the dependence of the cooling strength, $\chi$, in (a) and the frequency separation of $\omega_{\eta}$ and $\omega_{\xi}$ in (b) on the cooling rate, $<P>_{\xi\dot{\xi}+\eta\dot{\eta}}$. (a) Intermittently increasing the cooling strength $\chi$ during a single cooling process for a fixed electric field strength ($\theta = \frac{4\pi}{32}$). Each dip corresponds to an abrupt increase in the value of $\chi$. At $t=0$, the cooling process starts with $\chi = 10^{7}~\text{s}/\text{m}^{2}$. Beginning with $t=3$ ms, $\chi$ is increased every 1 ms by a factor of ten, ending with $10^{12}~\text{s}/\text{m}^{2}$. (b) Average energy after feedback cooling versus $\theta$ showing the dependence of the frequency separation between $\omega_{\eta}$ and $\omega_{\xi}$ on the cooling rate $<P>_{\xi\dot{\xi}+\eta\dot{\eta}}$. The points are averages of 1000 calculated energies following feedback cooling for a fixed simulation time of 80 ms and cooling strength $\chi = 10^{7}~\text{s}/\text{m}^{2}$.  }
\end{figure} 

Figure \hyperref[fig:histograms]{3(d)} shows the final energies plateauing near 5 K. This is a consequence of the simulation time used of 80 ms and not a limit to further cooling. The limit is set only by the accuracy of the simulations. What delays further energy reduction are the decreasing values of the $A_{\pm}(t)$ in the cooling rate $<P>_{\xi\dot{\xi}+\eta\dot{\eta}}$. To circumvent this delay one may intermittently increase the cooling strength $\chi$ to achieve more rapid cooling as shown in Fig. \hyperref[fig:power_depend]{4(a)}. As an example, for the nanoparticle considered in this paper, an occupation number $n=1$ corresponds to a temperature on the order of $T = \hbar\omega/k_{B}=16.7~\mu$K for $\omega = 2.19$ MHz. Setting the simulation accuracy to $\sim10^{-10}$ K, the particle is able to reach a temperature of $\sim10^{-9}$ K by employing the same method as that in Fig. \hyperref[fig:power_depend]{4(a)}. These classical calculations thus show that parametric feedback cooling is a suitable method for approaching the quantum regime. The dynamics and fundamental limits at lower temperatures will require a full quantum analysis and will be addressed in a future report.

Also affecting the cooling rate is the frequency separation between $\omega_{\xi}$ and $\omega_{\eta}$. A slight difference in frequency will allow cooling, but the rate is much larger when the frequency difference is larger. In Fig. \hyperref[fig:power_depend]{4(b)} the final average energy of 1000 randomly initialized cooling runs, $<\varepsilon>$, is plotted versus $\theta$ with each run having a fixed simulation time of 80 ms. For $\theta \approx 0$ ($\omega_{\eta}\approx\omega_{\xi}$) the average final energy is $\sim300$ K, similar to the final temperature when feedback cooling using linear polarization. As $\theta$ increases ($\omega_{\eta}>\omega_{\xi}$), the cooling proceeds more quickly, as evidenced by the average final energy decreasing. The rate plateaus near $\theta=\frac{4\pi}{32}$ where the competing heating terms in Eqs. (\hyperref[eq29]{29}) and (\hyperref[eq30]{30}) become negligible. 

\subsection{\label{exp_sig}Experimental Signatures}

\begin{figure}[h]  
  \centering
    \hspace*{-0.5cm}\includegraphics[width=0.42\textwidth]{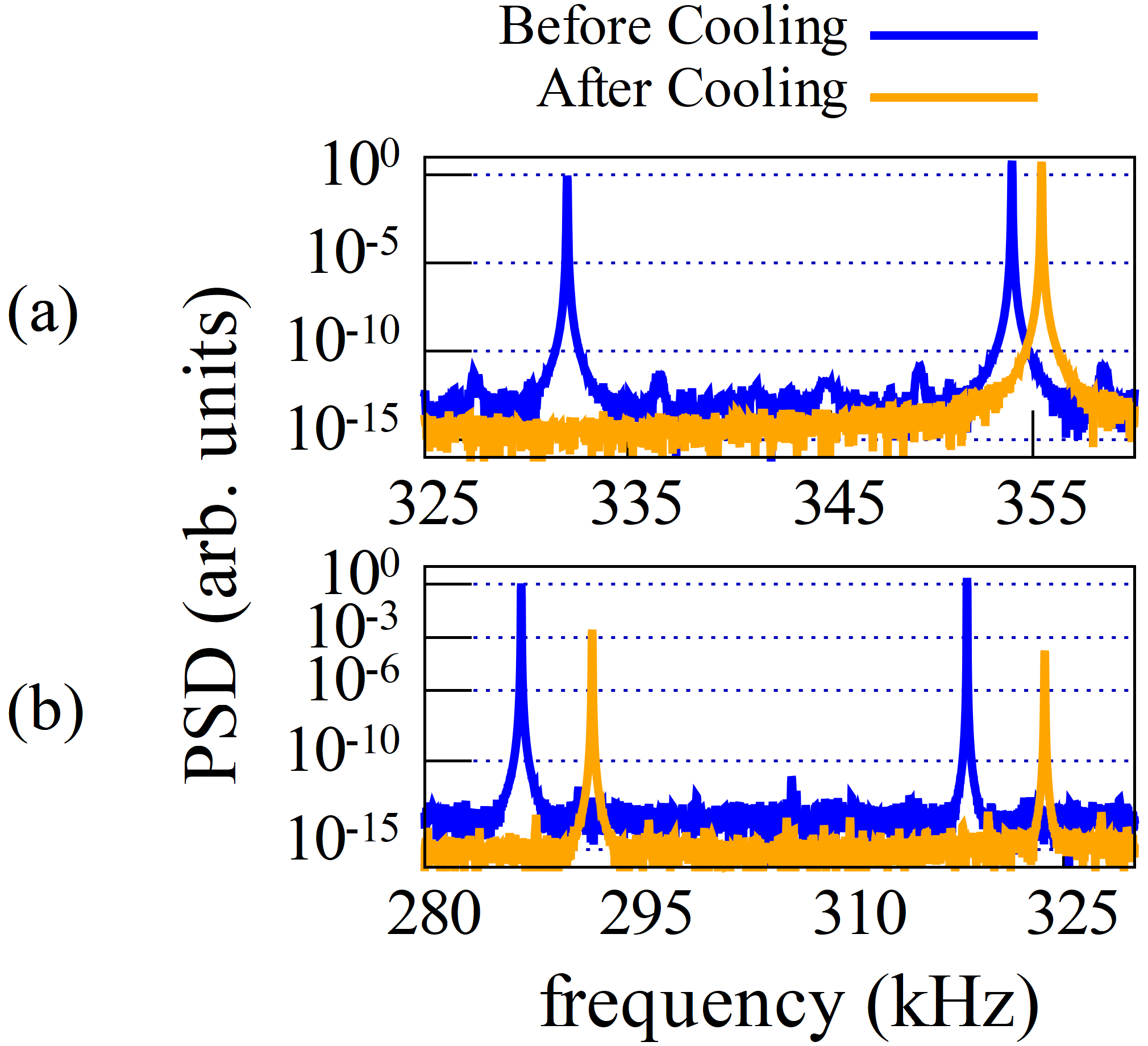} \label{fig:psds} 
    \caption{Power spectral densities of a $p_{y}$ measurement before and after feedback cooling for (a) linear polarization and (b) elliptical polarization. Feedback cooling using linear polarization eliminates one peak, shifting the remaining peak to a normal mode frequency and reducing the motion to pure precession. Feedback cooling under elliptical polarization reduces both peaks in magnitude, and shifts them toward the normal mode frequencies found in the small angle approximation.}
\end{figure}

What is actually measured in the laboratory is the power spectral density (PSD) of the signal. Figure \hyperref[fig:psds]{5(a)(b)} shows the PSD of a $p_y$ measurement before and after cooling the particle using linear and elliptical polarization. Before cooling, two peaks are seen identifying the existence of two rotational motions at different frequencies; the libration and precession motions discussed in Sec. \ref{Model}. As the particle is cooled using linear polarization, both peaks converge to a normal mode frequency $\omega_{\pm}$ with the larger peak reducing to a non-zero value and the smaller peak decreasing to zero. Introducing elliptical polarization allows both modes to be cooled fully. In this case, the two initial peaks each converge to a normal mode frequency with the magnitudes of both peaks decreasing to zero.

\section{\label{Discussion}Discussion of heating and noise}   
The above analysis has shown that it is theoretically possible to cool the librational motion through parametric feedback cooling within a classical approximation that does not include sources of noise or heating. However, real experiments will encounter unavoidable shot noise, gas collisions, measurement uncertainty, and quantum limits. 

The effects of a non-zero measurement uncertainty has been addressed in \cite{Zhong2017} showing that inefficient feedback sets a lower bound on the occupation number after cooling for a fixed cooling power. This noise source becomes important for a nanoparticle in a low occupation state $n\sim 50$ ($T\sim1$ mK). The spin about the symmetry axis of a nanodumbbell limits the energy of the particle's accessible DOF in the 1-100 K range after parametric feedback cooling with linear polarization (Fig. \hyperref[fig:histograms]{3(c)}), leaving imperfect feedback to be a negligable effect. For elliptical polarization, librational cooling is expected to be affected similarly to that found in \cite{Zhong2017} with a lower bound on the occupation number. A full quantum treatment, to be performed in the future, will test this hypothesis.

One may think it possible that gas collisions could induce an asymmetry between the librational coordinates which would allow further cooling. To test this hypothesis the effects of shot noise and gas collisions in our simulations were included for linearly polarized light. Laser shot noise was included using the methods of \cite{Zhong2017}. Gas collisions were considered as the Langevin type, $\dot{\pi_{i}}=-\Gamma_{i}\pi_{i}+ \zeta(t)$, with $\pi_{i} = (\dot{\alpha}$, $\dot{\beta}$, $\dot{\gamma}$), $\Gamma_{i} = \tau_{i}/I_{i}\Omega_{i}$ the damping rate ($\Gamma_{\alpha}=\Gamma_{\beta}$) \cite{Ahn2018}, and $\zeta(t)$ stochastic noise. Simulations were performed for three different pressures $P=$760, $10^{-3}$, and $10^{-7}$ Torr. 

For $P=$760 Torr, the nanoparticle is unable to be cooled. The final energies conform to a Maxwell-Boltzmann distribution as it thermalizes with the surrounding gas at 300 K. Here, increasing the cooling strength $\chi$, with hope to overcome energy exchange with the gas, heats the particle as its motion is more Brownian than periodic. 

For $P=10^{-3}$, and $10^{-7}$ Torr, the main results of Secs. \hyperref[Model]{II} and \ref{Linear_cooll} hold, with final energy distributions and PSD's similar to that of Fig. \hyperref[fig:histograms]{3(c)} and Fig. \hyperref[fig:psds]{5(a)}, respectively. The simulations reveal that gas collisions and photon scattering do not change the general conclusions of this paper in the classical limit. The effects of laser shot noise and gas collisions while cooling using elliptically polarized light are expected to limit the lowest occupation number attainable for a fixed cooling rate and will be studied in a future report. 

\section{\label{Conclusion}Conclusion}
We have theoretically studied the rotational dynamics of an optically trapped nanodumbell with and without parametric feedback cooling. A relatively simple model describing the motions in a small angle approximation has also been provided. The nanoparticle oscillates about the polarization axis as a superposition of two modes resulting in a combination of libration and precession motions. The librational motion is due to the laser field's potential while precession arises from the non-zero spin of the nanoparticle about its symmetry axis. The equations of motion describing the location of the tip of the nanoparticle in the small angle approximation are seen to have the same form as a charged particle in a harmonic oscillator potential and a static magnetic field. 

The effect of parametric feedback cooling using a linearly polarized beam is to remove one of the two modes, resulting in pure precession. In this geometry, it is not possible to extract energy from more than two degrees of freedom and not possible to cool to the quantum regime even when information about both librational modes is available. Evidence of these dynamics may be found in the power spectral density with two peaks converging toward normal mode frequencies during the cooling process, with the smaller of the two peaks' magnitude reducing to zero. 

Using a potential energy that sets different frequencies of libration allows cooling to much lower energies when information about both librational modes are available, theoretically approaching the quantum regime in this classical analysis. The setup for cooling may be acheived experimentally by using elliptical polarization or using two perpendicular laser beams incident on the nanoparticle and feeding back both coordinate frequencies.  If a single librational coordinate frequency is used in the feedback, the particle will ultimately heat. The rate of cooling is largely determined by the cooling strength and the separation between the two librational frequencies. In this case, the power spectral density will show two peaks converging toward the two normal mode frequencies with both magnitudes decreasing to zero over time. 

After submission of this manuscript, Ref. \cite{Ge2018} was published proposing a method for cooling the librational motion of an ellipsoidal nanodiamond utilizing the intrinsic magnetic dipole moment of the NV center. This method is only useful for nanodiamonds, but may have potential to cool into the quantum regime. However, Ref. \cite{Ge2018} did not include the rotation about the symmetry axis. Therefore, it is difficult to ascertain whether the symmetry discussed in Sec. \ref{Cooling} would be relevant. In particular, it is uncertain whether this method would allow cooling of the rotation about the symmetry axis and/or cooling of more than one librational mode.

\begin{acknowledgments}

We would like to acknowledge Prof. Tongcang Li, Jaehoon Bang, Jonghoon Ahn, and Zhujing Xu for their fruitful insights and discussions. This work was supported by the Office of Naval Research (ONR) Basic Research Challenge (BRC) under Grant No. N00014-18-1-2371.

\end{acknowledgments}

\appendix
\section{\label{appendixA}Convention and Dynamics}

The Euler angles in the $z$-$y'$-$z''$ convention have been used, which, for the sake of clarity, gives the rotation matrix $\overset{\text{\tiny$\leftrightarrow$}}{R}$ as
\begin{equation}
\overset{\text{\tiny$\leftrightarrow$}}{R} = \overset{\text{\tiny$\leftrightarrow$}}{R}_{z''}\overset{\text{\tiny$\leftrightarrow$}}{R}_{y'}\overset{\text{\tiny$\leftrightarrow$}}{R}_{z},
\end{equation}
and 
\begin{align}
\overset{\text{\tiny$\leftrightarrow$}}{R}_{z} &=  \begin{pmatrix}  \cos\alpha & \sin\alpha & 0 \\
																								 -\sin\alpha & \cos\alpha & 0 \\
																									0 & 0 & 1 \end{pmatrix} , \\
\overset{\text{\tiny$\leftrightarrow$}}{R}_{y'} &=  \begin{pmatrix}  \cos\beta & 0 &  -\sin\beta \\
																								 0 & 1 & 0 \\
																								\sin\beta & 0 &  \cos\beta  \end{pmatrix}, \\
\overset{\text{\tiny$\leftrightarrow$}}{R}_{z''} &=	\begin{pmatrix}  \cos\gamma & \sin\gamma & 0 \\
																								 -\sin\gamma & \cos\gamma& 0 \\
																									0 & 0 & 1 \end{pmatrix}. 
\end{align}
The full equations of motion for $\alpha , \beta , \gamma$ are found through the Lagrangian with kinetic and potential energies 
\begin{align} 
K &= \frac{1}{2}I_{x}(\omega_{1}^2+\omega_{2}^2) + \frac{1}{2}I_{z}\omega_{3}^2 , \\
U &= -\frac{1}{4} \vec{p}\cdot \vec{E}_{inc} , \label{eqA6 }
 \end{align}
where $\vec{p}=\overset{\text{\tiny$\leftrightarrow$}}{R}^\dagger \overset{\text{\tiny$\leftrightarrow$}}{\alpha}_{0}  \overset{\text{\tiny$\leftrightarrow$}}{R} \vec{E}_{inc} $ is the polarization vector, $\overset{\text{\tiny$\leftrightarrow$}}{\alpha}_{0}$ is the diagonal polarizability matrix, and the body frame angular velocities are given by
\begin{align} 
\omega_{1} &=  \dot{\beta}\sin(\gamma) - \dot{\alpha}\sin(\beta)\cos(\gamma) ,  \\
\omega_{2} &=  \dot{\beta}\cos(\gamma) + \dot{\alpha}\sin(\beta)\sin(\gamma) , \\
\omega_{3} &=  \dot{\alpha}\cos(\beta) + \dot{\gamma}  = const. \label{eqA9}
\end{align}
Due to the particle symmetry, $\omega_{3}$ is a constant of the motion. The equations of motion for the three angles are
\begin{align} 
\ddot{\alpha} &= -2 \dot{\alpha} \dot{\beta} \cot(\beta) + \dot{\beta}\csc(\beta) \frac{I_{z}}{ I_{x}} \omega_{3} -\frac{1}{I_{x}\sin^{2}(\beta)}\Big(\frac{\partial U}{\partial \alpha}\Big) , \\
\ddot{\beta} &= \sin(\beta)\left(  \dot{\alpha}^2 \cos(\beta) - \dot{\alpha} \frac{I_{z}}{ I_{x}} \omega_{3} \right) -\frac{1}{I_{x}}\Big(\frac{\partial U}{\partial \beta}\Big) , \\
\dot{\gamma} &=  \omega_{3}-\dot{\alpha}\cos(\beta) . \label{eqA12}
 \end{align}
To evaluate Eq. (\ref{eqA6 }), we consider the nanodumbbell in the dipole limit $\lambda>>R$ with the center of mass fixed at the origin so that the electric field has no spatial dependence. For elliptical polarization, $\vec{E}_{inc}=E_{0}<\cos\theta,i\sin\theta,0>$, the potential is
\begin{align} \begin{split}
U = - &\frac{E^{2}_{0}}{4}\bigg[\alpha_{x}+ \big( \alpha_{z}-\alpha_{x} \big)\sin^{2}\beta\\
&\times\Big( \cos^{2}\theta\cos^{2}\alpha + \sin^{2}\theta\sin^{2}\alpha \Big) \bigg] ,
\end{split} \end{align}
and the potential for linear polarization is $U(\theta=0)$. The analysis in the main paper excludes the constant term in the potential energy, as it does not affect the librational and rotational dynamics.

\section{\label{appendixB}Axial Displacement in the Laser Field}
To determine the approximate size of translational displacement in the axial direction, consider a laser with intensity $I_{0}$, power $P_{0}$, and wavevector $\vec{k}=\frac{2\pi}{\lambda}\hat{z}$ incident on a dumbbell of radius $R$ and polarizability $\alpha_{0} = 4\pi \epsilon_{0}R^{3} \overline{\alpha}$ with $\epsilon_{0}$ the permittivity of free space. The average force on the nanoparticle due to momentum transfer from the beam in the axial direction is
\begin{align}
F_{z} &= \Big( \frac{I_{0}\lambda}{hc} \Big) \int_{\Omega} \frac{d\sigma}{d\Omega}\big(\Delta \vec{p}\cdot \hat{z}\big) d\Omega \\
&=  \Big( \frac{4}{3} \Big)  \Big( \frac{P_{0}\text{NA}^{2}}{c} \Big)  \Big( \frac{2\pi R}{\lambda} \Big)^{6} \overline{\alpha}^{2} ,
\end{align}
where $h$ is Planck's constant, $c$ is the speed of light, $\Delta \vec{p} = \frac{h}{\lambda}\Big[ \big(1-\cos(\theta)\big)\hat{z} +\sin(\theta)\hat{\rho}\Big]$ is the momentum transfer function with $\theta$ the angle with respect to the z-axis, and $\frac{d\sigma}{d\Omega}=k^{4}R^{6}\overline{\alpha}^{2}\big( 1-\cos^{2}(\theta)\sin^{2}(\phi)\big)$ is the differential scattering cross section for a particle in the point dipole limit. The size of the displacement $z_{d}$ is estimated by looking at the equations of motion to first order with the nanoparticle in its equilibrium position 
\begin{align}
m\ddot{z} &= 0 = -m\omega^{2}_{z}z_{d}  +F_{z} , \label{B3}
\end{align} 
where $m\omega^{2}_{z} = \Big( 2\alpha_{0} (\text{NA})^{6} \pi^{3}/(c\epsilon_{0}\lambda^{4}) \Big)P_{0}$ \cite{PhysRevLett.109.103603}. Solving for $z_{d}$ in Eq. (\ref{B3}) recovers the expression found in Eq. (\ref{eq19}),
\begin{equation}
\frac{z_{d}}{z_{R}} = \Big(\frac{32\overline{\alpha}}{3}\Big)\Big(\frac{\pi R}{\lambda}\Big)^{3}\Big(\frac{1}{\text{NA}}\Big)^{2}.
\end{equation}

\section{\label{appendixC}Parametric Feedback Cooling Under Elliptical Polarization}
The normal modes of Eqs. \hyperref[eq27]{(27)} \hyperref[eq28]{(28)} are 
\begin{align}
\xi(t) &= A_{+} \cos(\omega_{+} t +\delta_{+}) + A_{-} \cos(\omega_{-} t +\delta_{-}) , \\
\eta(t) &= A_{+}\kappa_2 \sin(\omega_{+} t +\delta_{+}) - A_{-}\kappa_1 \sin(\omega_{-} t +\delta_{-}), 
\end{align}
with $\omega_{\pm}=\frac{1}{\sqrt{2}}\Big( \omega^{2}_{\xi}+\omega^{2}_{\eta}+\omega^{2}_{c} \pm Q\Big)^{\frac{1}{2}}$, $Q=\sqrt{4\omega^{2}_{\eta}\omega^{2}_{c}+\left( \omega^{2}_{c}+\omega^{2}_{\xi}-\omega^{2}_{\eta} \right)^{2}}$, $\kappa_{1}=\left(2\omega_{+}\omega_{c}\right)/\left(Q+\omega^{2}_{c}+\omega^{2}_{\xi}-\omega^{2}_{\eta}\right)$, $\kappa_{2}=\left(2\omega_{-}\omega_{c}\right)/ \left(Q-\omega^{2}_{c}-\omega^{2}_{\xi}+\omega^{2}_{\eta} \right)$. The $\kappa_{i}$ ($i=1,2$) have the following relations $\kappa^{2}_{1}\geq1$ ,  $\kappa^{2}_{2}\leq1$, $\kappa_{i}(\theta=0)=1$. The relations are important when considering the cooling rate when feeding back twice of both coordinate's frequencies. The small angle approximation equations of motion under cooling become
 \begin{align}
\ddot{\xi} &= -\omega^{2}_{\xi}\Big(1+\chi R^{2}(\xi\dot{\xi}+\eta\dot{\eta}) \Big)\xi - \omega_{c}\dot{\eta} , \\
\ddot{\eta}  &= -\omega^{2}_{\eta}\Big(1+\chi R^{2}(\xi\dot{\xi}+\eta\dot{\eta}) \Big)\eta + \omega_{c}\dot{\xi}.
\end{align}
The cooling rate is the addition of Eqs. (\hyperref[eq29]{29}) (\hyperref[eq30]{30}) ,
\begin{equation} \begin{split}
<P>_{\xi\dot{\xi}+\eta\dot{\eta}} &= <P>_{\xi\dot{\xi}}+<P>_{\eta\dot{\eta}} \\
&= -\Big[ A_{+}(t)\Big]^{4}\Big(z_{1}-y_{1}\Big) -\Big[ A_{-}(t)\Big]^{4}\Big(y_{2}-z_{2}\Big) \\
&\quad- \Big[ A_{+}(t)A_{-}(t)\Big]^{2}\Big(y_{3}+z_{3}\Big).   \label{eqC5}\end{split}
\end{equation}
The coefficients in Eq. (\ref{eqC5}) are constant for fixed electric field strengths and are as follows,
\begin{align}
\Big(z_{1}-y_{1}\Big) &= \bigg( \frac{I_{x}\omega^{2}_{+}\chi R^{2}}{4} \bigg)\big(\kappa^{2}_{1}-1 \big)\big(\omega^{2}_{\eta}\kappa^{2}_{1}-\omega^{2}_{\xi} \big) \geq0 , \\
\Big(y_{2}-z_{2}\Big) &= \bigg( \frac{I_{x}\omega^{2}_{-}\chi R^{2}}{4} \bigg)\big(1-\kappa^{2}_{2} \big)\big(\omega^{2}_{\xi}-\omega^{2}_{\eta}\kappa^{2}_{2} \big) \geq0 , \label{eqC7} \\
\Big(y_{3}+z_{3}\Big) &= \bigg(\frac{I_{x}\omega^{2}_{\xi}\chi R^{2}}{2}\bigg) \bigg( 4\omega^{2}_{\xi}+\omega^{2}_{c} \bigg) >0 ,
\end{align}
which leads to complete cooling for the conditions described in this paper ,  $\omega^{2}_{\eta}>\omega^{2}_{\xi}\gg \omega^{2}_{c}$. Only for very large values of $\omega_{c}$ ($\omega_{c}/\omega \sim 10^{5}$) is $\Big(y_{2}-z_{2}\Big)<0$.

\bibliographystyle{apsrev4-1}
\bibliography{Levitated_Optomechanics}

\end{document}